\documentclass[aip,pop,reprint,numerical,twocolumn,superscriptaddress]{revtex4}
\usepackage{amssymb, amsmath,framed}

\usepackage{bm}
\usepackage[colorlinks=true,linkcolor=blue]{hyperref}
\expandafter\ifx\csname package@font\endcsname\relax\else
 \expandafter\expandafter
 \expandafter\usepackage
 \expandafter\expandafter
 \expandafter{\csname package@font\endcsname}
\fi
\def\bq{\begin{equation}}
\def\eq{\end{equation}}
\def\bqy{\begin{eqnarray}}
\def\eqy{\end{eqnarray}}

\def\de{\delta}

\def\ep{\epsilon}

\def\p{\partial}

\def\rh{\rho}

\def\p{\partial}

\def\R{\mathbb{R}}

\def\calb{\mathcal{B}}

\def\calj{\mathcal{J}}

\def\call{\mathcal{L}}
\def\calm{\mathcal{M}}

\def\calp{\mathcal{P}}

\def\cals{\mathcal{S}}

\def\calv{\mathcal{V}}

 \def\q#1#2{q^{\hspace{1pt}#1}_{\,,\hspace{1pt}#2}}
 \def\a#1#2{a^{\hspace{1pt}#1}_{\,,\hspace{1pt}#2}}

\begin{document}

\title{The action principle for generalized fluid motion including gyroviscosity}

\author{M.~Lingam}
\email{manasvi@physics.utexas.edu}
\affiliation{Department of Physics and Institute for Fusion Studies, The University of Texas at Austin, Austin, TX 78712}
\author{P.~J.~Morrison}
\email{morrison@physics.utexas.edu}
\affiliation{Department of Physics and Institute for Fusion Studies, The University of Texas at Austin, Austin, TX 78712}

\date{August 2, 2014}

\begin{abstract}

A general set of fluid equations that allow for energy-conserving momentum transport by gyroscopic motion of fluid elements is obtained.  The equations are produced by a class of action principles that  yield a large subset of the known fluid and magnetofluid models, including gyroviscosity.   Analysis of the action principle  yields broad, model-independent results regarding the conservation laws of energy and  linear and angular momenta. The  formalism is illustrated by  studying   fluid models with intrinsic angular momentum that may appear in the contexts of condensed matter, biological, and other areas of  physics.

\end{abstract}

\maketitle


\section{Introduction}
\label{sec:intro}

Fluid models have been  effectively used to describe  a vast range of physical phenomena,   from  microscopic  to  macroscopic scales,  in fields as diverse as plasma physics, condensed matter physics, oceanography, atmospheric science, geophysics,  and astrophysics. Often these models are obtained by using phenomenological methods or other modes of reasoning.  In contrast, in this paper, we present a framework for extracting fluid  equations of motion from a general  action principle.  The action principle produces a general class of fluid equations  that include  the possibility of transporting momentum by gyroscopic motion by means of a  gyroviscosity tensor that conserves energy. 

We work with the action as our central object, as it represents an instantly recognizable and transparent method of deriving dynamical equations for models. The action formalism dates back to the original pioneering work by Lagrange \cite{Lagrange}, which  was extended by many illustrious scientists  (e.g., \cite{clebsch57,clebsch59,helmholtz58,hankel,kirchhoff76,serrin59,newcomb62,newcomb73})  in the 19th and 20th centuries. The action formalism is closely tied to the Hamiltonian approach, which involves the use of noncanonical Poisson brackets, and this field was reinvigorated following the crucial work of \cite{morrison80}. A summary of this approach can be found in \cite{morrison82,morrison98,morrison05,morrison06}. As the Hamiltonian and action principle approaches are mutually complementary, with a close connection between the two, we shall refer to them henceforth as the Hamiltonian and Action Principle (HAP) formalism.

The use of action principles is  ubiquitous, as it is a basic tool in general relativity, high energy physics and condensed matter physics.  However, for fluid, plasma, and other matter dynamics one must guarantee the existence of a set of Eulerian variables, as we shall see.  In the context of plasma physics, the HAP formalism has been employed in  magnetohydrodynamics (MHD) \cite{morrison80}, the Vlasov description (e.g., \cite{morrison82,morrison13}), and the BBGKY hierarchy \cite{MMW84} as all these models possess an underlying Hamiltonian structure. There is yet another reason to employ such an approach, as it eliminates  ``spurious" dissipation, i.e.,  several models are often claimed to be energy conserving, even when they aren't. For a discussion of the same, we refer the reader to \cite{HKM85,HKM86, sc05,tronci, kimura}. The HAP formalism enables us to construct and build non-dissipative models from first principles via a transparent procedure.

The general procedure for building fluid action principles is described in the companion works \cite{morrison09,ling}. We note that the HAP approach enables us, amongst other virtues, to construct  versions of MHD \cite{amp0,amp1,amp2a}, reduced fluid models  \cite{strauss76,MH84,MM84,HKM85,KPS94,KK04,TMG07,TMWG08,WMH04,WHM09,TM00,tassi10}), gyrofluids \cite{CC53,kaufman60,newcomb72,newcomb73,newcomb83,newcomb90}, the Oldroyd-B fluid \cite{old50,OP03}, nematic fluids \cite{st05}, and to explain the origin of the gyromap, a tool introduced in \cite{MCT84} and used in previous derivations of reduced fluid models  \cite{HHM86,HHM87,ICTC11}.
   
The remainder of the paper is organized as follows.  In Sec.~\ref{ActPrin}, we describe the underlying physical and mathematical principles in building up actions. In Sec.~\ref{APapproach}, the basic approach is highlighted, and then applied to build the gyroviscous fluid -  a general class of fluid and magnetofluid models. The resultant equation of motion is analyzed in Sec.~\ref{Analysis}, and some general comments about conservation laws are presented.  In 
Sec.~\ref{rotMHD} illustrate the methodology by showing how to produce a  fluid model that includes intrinsic angular momentum, and  highlight potential systems where such models might be of interest.  Lastly, in Sec.~\ref{Conclusion}, we round up our analysis by presenting avenues for future work.


 \section{The action principle and the Lagrangian coordinates}
 \label{ActPrin}
 The action principle for particles involves a standard prescription, which we shall not describe in detail here. First, a set of generalized coordinates, denoted by $q_k(t)$ are chosen, where $k$ runs over all possible degrees of freedom. The action principle is given by
\bq
S[q] = \int^{t_1}_{t_0} \! dt\,L\left(q,\dot{q},t\right)\,,
\eq
where $L=T-V$ represents the Lagrangian, with the kinetic and potential energies denoted by $T$ and $V$ respectively. We note that $S$ serves as a ``functional", i.e. it has a domain of functions and a range of real numbers. Hamilton's principle of least action states that the equations of motion can be found by extremizing the action, i.e. we require $\delta S[q]/\delta q^k = 0$. The functional derivative is defined via
 \bq
\de S[q;\de q]=
\left.\frac{d S[q +\ep \de q]}{d \ep}\right|_{\ep =0}
=: \left\langle \frac{\de S[q]}{\de q^{i}},\de q^{i}\right\rangle\,.
\eq
It is natural to seek a generalization of the above procedure for continuous media. As such, the discrete label $k$ is replaced by a continuous one, which we denote by $a$. As a result $q$ is now a function of $a$ and $t$, and tracks the location of a fluid particle labelled by $a$. Two associated quantities which shall recur in this paper are the deformation matrix, $\p q^i/\p a^j=:\q ij$ and the corresponding determinant, the Jacobian, $\mathcal{J}:= \det(\q ij)$. A  volume element evolves in time via
\bq
d^3q=\mathcal{J} d^3a\,,
\label{vol}
\eq 
and an area element is governed by
\bq
(d^2q)_i=  \mathcal{J} \a ji \, (d^2a)_j\,,
\label{area}
\eq 
where $\mathcal{J} \a ji$ is the transpose of the cofactor matrix of $\q ji$. From the definitions of $q$, the Jacobian, the deformation matrix, etc.\  one can derive a host of identities which we shall not reproduce here; instead we refer the reader to  \cite{serrin59,morrison98}.


\subsection{The Lagrangian and Eulerian pictures} 
\label{ssec:AOLE}
The Lagrangian picture, as we have seen, is based on  the Lagrangian coordinate $q$ for a fluid element, which is solely labelled by $a$. However, a fluid particle can also carry other properties with it. It may be endowed with some mass density, entropy or a magnetic field in the case of magnetofluid models. These quantities are attached to the fluid particle, and are consequently dependent on $a$ alone. We refer to them as {\it attributes},  since they are intrinsic to the fluid. As they depend solely on $a$, these serve as Lagrangian constants of motion. The subscript $0$ is used to describe the attributes, in contrast to their Eulerian counterparts.

The Eulerian picture is used because it allows for an easy description in terms of observable parameters. All Eulerian fields depend on the position $r:=(x^1,x^2,x^3)$ and $t$, which can both be measured in the laboratory. As a result, we refer to    these fields as {\it observables}. We describe the Lagrange-Euler maps, which allow us to move from one description to another.

First, let us consider the velocity field $v(r,t)$.  A measurement of $v$ corresponds to determining the velocity of the fluid element at a location $r$ and time $t$. From the Lagrangian picture, this must also equal $\dot{q}(a,t)$, since we wish to preserve an equivalence between these two frameworks. Thus, we see that $\dot{q}(a,t)=v(r,t)$, where the overdot indicates that the time derivative is obtained at fixed $a$.  This relation is incomplete because $a$ hasn't been specified. However, we note that the fluid element is at $r$ in the Eulerian picture, and at $q$ in the Lagrangian one. Thus, we see that $r=q(a,t)$, which implies that $a=q^{-1}(r,t)=:a(r,t)$.  As a result, the Eulerian velocity field is given by
\bq
v(r,t) =\left.\dot{q}(a,t)\right|_{a=a(r,t)}\,.
\eq
The above expression is an example of the Lagrange to Euler map that allows us to move back and forth between two different pictures. 

Now, we consider the attributes defined earlier, which are carried by the fluid. A fluid may have a certain entropy associated with it, which we denote by $s_0$. In an ideal fluid, we expect the entropy to be conserved along the fluid element. In other words, the Eulerian entropy $s(r,t)$ must remain constant throughout, implying that $s=s_0$. This amounts to $s$ behaving as a zero-form. We denote all attributes that obey this property by $\cals_0^\alpha$ and the corresponding observables by $\cals^\alpha$, where $\alpha$ runs over all such fields. Apart from entropy, the magnetic stream function $\psi$ for 2D gyroviscous MHD \cite{amp2a,ling} also obeys this property.

Next, we can consider attributes which obey a conservation law similar to the density. Let us denote the attribute by $\rho_0(a)$ and the observable by $\rh(r,t)$. Mass conservation in a given volume dictates that \ $\rho(r,t)d^3r=\rho_0d^3a$.  By using (\ref{vol}) we obtain $\rho_0=\rho \calj$. As a result, we have found a prescription for $\rho$, geometrically interpretable as a three-form. Other attributes (and their corresponding observables), such as the entropy density, may also obey a similar conservation law. We denote them by $\calp_0^\beta$ and $\calp^\beta$ respectively. 

A natural extension involves the magnetic field $B_0(a)$ carried by a given fluid element, which satisfies the frozen flux constraint. Mathematically, this amounts to $B\cdot d^2r=B_0\cdot d^2a$, and from (\ref{area}) we obtain $\calj B^i=\q ij  \,B_0^j$.  In other words, the magnetic field $B$ can be interpreted as a two-form \cite{deschamps81,dual}. As before, we generalize this to include other fields that satisfy frozen flux constraints, and denote the attribute-observable pairs by $\calb_{0i}^\gamma$ and $\calb_i^\gamma$,  respectively. 

In each of the above expressions, we see that there is a mismatch since the label $a$ is present in the attributes. To complete the Lagrange-Euler maps, we evaluate the attributes at $a=q^{-1}(r,t)=:a(r,t)$. As a result, one can now construct observables once the attributes and the field $q(a,t)$ are known. 

There exists a more intuitive way to represent the Lagrange to Euler map  in terms of  integrals. Let us suppose that we are given the attribute-observable relationships described above. In order to move from the Lagrangian description to the Eulerian one, we need to `pluck out' the fluid element that happens to be at the Eulerian observation point $r$ at time $t$. This is accomplished via the delta function $\delta(r-q(a,t))$. For instance, the three forms described above are obtained via
\bqy
\calp^\beta({r},t)&=&\int_D \!d^3a
\, \calp_0^\beta(a) \, \de\left({r}-{q}\left(a,t\right)\right)
\nonumber\\
&=&\left. \frac{\calp^\beta_0}{\mathcal{J}}\right|_{a=a({r},t)}\,.
\label{rhoEu3D}
\eqy
We will introduce the  momentum density, $M^c=(M^c_1,M^c_2,M^c_3)$, which is related to the Lagrangian canonical momentum through the expression
\bqy
M^c(r,t)&=&\int_D \!d^3a \,
{\Pi}(a,t) \, \de\left({r}-{q}(a,t)\right) 
\nonumber\\
&=& \left.
\frac{\Pi(a,t)}{\mathcal{J}}
\right|_{a=a(r,t)}\,.
\label{Mcan3D}
\eqy
The superscript `$c$' indicates that the momentum density constructed is the canonical one. For   MHD,  $\Pi(a,t)=(\Pi_1,\Pi_2,\Pi_3)=\rh_0 \dot{q}$.  In general, note that $\Pi(a,t)$ can be found from the Lagrangian through $\Pi(a,t) = {\de L}/{\de \dot{q} }$ and is not always equal to  $\rh_0 \dot{q}$.  Similarly, one can construct equivalent integral relations for  $\cals^\alpha$ and $\calb_i^\gamma$,  respectively. 

Hitherto, we have introduced Eulerian observables that behave geometrically as zero, two,  and three-forms,  respectively. Observables that behave as one-forms were not included in our description. One reason for this stems from the fact that, in three dimensions, they amount to the Hodge dual of the 2-forms and might lead to over specification (see, e.g., \cite{thiffeaultNYC}).  Furthermore, such quantities do not usually appear in the context of fluids and magnetofluids. Hence, we shall not consider such quantities in this paper, although they can be incorporated without any difficulty. 


\section{Action principle for the general gyroviscous fluid}
\label{APapproach}
In this section, we provide a brief summary of the general methodology advocated in  \cite{morrison09} for constructing action principles for fluid and magnetofluid models and obtain the gyroviscous fluid action. The advantages of this approach are manifold, and we shall refer the reader to \cite{morrison09,ling} for a discussion of the same. Next, we describe how we build our model and obtain the corresponding equation of motion, thereby proving the Eulerian closure principle along the way.

\subsection{The general action}
\label{process}
 
First, we choose the domain  $D\subset \R^{3}$.  We also assume the existence of the Lagrangian coordinate $q\colon D\rightarrow D$, which is a well behaved function that is sufficiently smooth, invertible, etc. Next, we specify our set of observables, which are fully determined by the attributes and $q$. In our case, the set corresponds to $\mathfrak{E}=\{v, \cals^\alpha,\calp^\beta,\calb^\gamma\}$. The last step involves the imposition of a {\it closure principle}, which is necessary for our model to be  `Eulerianizable.'\ \ Mathematically, this principle is implemented by demanding the action to be expressible fully in terms of our Eulerian observables. In other words, we require our action to be expressible as follows:
\bq
S[q]:= \int_D d^3a dt\,\call\left(q,\dot{q},\p q/\p a\right) =: \bar{S}\left[\mathfrak{E}\right]\,.
\label{Lagaction}
\eq
Now, we shall make one additional simplification: $\bar{S}=\int_D d^3r dt\,\bar{\call}$, where $\bar{\call}$,  can depend on the observables and their spatial and temporal derivatives of any order. However, for convenience we shall use the following ansatz for the Lagrangian density $\bar{\call}$: 
\bq
\label{Theaction}
\bar{S}=\int_D d^3r dt\,\bar{\call} \left(v, \cals^\alpha,\calp^\beta,\calb^\gamma, \nabla v, \nabla\cals^\alpha,\nabla\calp^\beta,\nabla\calb^\gamma \right)\,; 
\eq
i.e.,  that the action only involves the observables and their first-order spatial derivatives. Such a simplification is well-motivated since most of the widely used fluid and magnetofluid models possess this form. The generalization to higher derivatives is  straightforward.

To sum up, there are two simplifications employed in this model. Firstly, we assumed that our model does not have observables that are akin to one-forms and,  secondly, we chose the ansatz (\ref{Theaction}) for the action. In order to obtain the equation of motion, we must use Hamilton's principle to extremize the action (\ref{Lagaction}). We shall instead show how we can extremize the action (\ref{Theaction}), and how it leads to equations of motion that are purely Eulerian.

As a result, for our family of models, this amounts to proving the Eulerian closure principle, which states that a completely Eulerianizable action yields Eulerian equations of motion. We shall drop the overbar in the action and the Lagrangian density described in (\ref{Theaction}), to simplify the notation. For the same reason, we also drop the Greek indices $\alpha$, $\beta$ and $\gamma$ present in (\ref{Theaction}). 


\subsection{The Eulerian closure principle and equations of motion}
\label{ECP}

The variation of the action (\ref{Theaction}) yields
\bqy
\delta S &=& \int_D d^3r dt\, \Big(\frac{\delta S}{\delta v^k}\delta v^k + \frac{\delta S}{\delta \calb^k}\delta \calb^k 
\nonumber\\
&&\hspace{2 cm}+ \frac{\delta S}{\delta \calp}\delta \calp + \frac{\delta S}{\delta \cals}\delta \cals\Big).
\label{varact}
\eqy
However, we need to express the quantities $\delta \calb_k$, $\delta \cals$, etc in terms of $\delta q$ in order to derive the equation of motion.\cite{EP} We shall present this calculation in detail for $\delta \calp$, since it is the most convenient for illustrating the procedure. From (\ref{rhoEu3D}), we find that 
\bq
\delta \calp=-\int_D \!d^3a\, \calp_0(a) \, \p_k \de\left({r}-{q}\left(a,t\right)\right) \de q^k,
\eq
where the partial derivative is wrt $r$ now. Substituting this into the relevant component of (\ref{varact}) and integrating  by parts yields 
\bq
\int_D d^3r dt\, \frac{\delta S}{\delta \calp}\delta \calp = \int_D d^3a dt \calp_0 \left[\p_k\frac{\de S}{\de \calp}\right]_q \de q^k,
\label{varrhoact}
\eq
where the notation $\left[\p_k\frac{\de S}{\de \calp}\right]_q$ is   short-hand for 
\bq
\left[\p_k\frac{\de S}{\de \calp}\right]_q = \int_D d^3r  \de\left({r}-{q}\left(a,t\right)\right) \p_k\frac{\de S}{\de \calp}.
\label{Lagvar}
\eq
The above expression has a ready physical interpretation. We earlier mentioned that the observables can be generated from the corresponding attributes since the delta function allows us to `pluck out' the appropriate fluid element. Here, the converse relation is true: given an Eulerian field (expressed in terms of the observables), the delta function allows us to pluck out the Lagrangian counterpart. As a result, the quantity (\ref{Lagvar}) is fully Lagrangian, since the action is fully representable either in terms of $q$ and its derivatives, or in terms of the Eulerian observables. Hence,   the subscript $q$   denotes its Lagrangian nature.

 Let us now return to (\ref{varrhoact}) and  extremize the action.  This requires  everything appearing in front of $\de q^k$ must  vanish.  The contribution from the $\calp$ term is evidently 
\bq
\calp_0 \left[\p_k\frac{\de S}{\de \calp}\right]_q\,,
\eq
and since we know that the determinant $\calj \neq 0$, we can divide throughout by $\calj$.  Next, evaluating  this expression at the label $a=q^{-1}(r,t)$ and using (\ref{rhoEu3D}) gives the following Eulerian contribution form the $\calp$-term:
\bq
\calp \p_k\frac{\de S}{\de \calp}\, ,
\label{rhofinalvar}
\eq
where we have used the fact that $\left[\p_k({\de S}/{\de \calp})\right]_q$ evaluated at $a=q^{-1}(r,t)$ yields $\p_k ({\de S}/{\de \calp})$. This effectively amounts to taking the quantity $\p_k ({\de S}/{\de \calp})$ and Lagrangianizing it (expressing it in terms of $q$, its derivatives and the attributes) and then re-Eulerianizing it again (re-expressing in terms of the Eulerian fields). We can also derive the same relation, by using the approach outlined in \cite{FR60}. With  the notation employed in \cite{amp2a} where the Lagrangian variation $\de q$ is denoted by $\xi$ and the Eulerianized counterpart is denoted by $\eta$, the variation for $\calp$ takes  the form 
\bq
\de \calp = - \p_k \left(\calp \eta^k \right).
\eq
Substituting this into (\ref{varrhoact}), integrating by parts and separating out the algebraic expression in front of $\eta^k$, gives  the same result as (\ref{rhofinalvar}). 

Consider now the $\cals$-term.  Since $\cals = \cals_0$, when the RHS is evaluated at $a=q^{-1}(r,t)$,  the integral representation of  this amounts to
\bq
\cals = \int_D d^3a\, \cals_0 \calj \de\left({r}-{q}\left(a,t\right)\right)\,.
\label{entrop}
\eq
Again, with this term we can either carry out the approach outlined above, or use the equivalent approach described in \cite{FR60}.  Substituting  (\ref{entrop}) into the appropriate term in (\ref{varact}),   obtaining the Lagrangian expression, dividing throughout by $\calj$,  and Eulerianization gives
\bq
\cals \p_k\frac{\de S}{\de \cals} - \p_k\left(\cals \frac{\de S}{\de \cals} \right) \,. 
\label{entropfinalvar}
\eq

Next, we consider the variable $\calb$-term, which satisfies the relation
\bq
\calb^j = \int_D d^3a\, q^j_{,i} \calb_0^i \de\left({r}-{q}\left(a,t\right)\right).
\label{Bdesc}
\eq
Repeating the   procedure,  for this term gives 
\bq
\calb^j \p_k\frac{\de S}{\de \calb^j} - \p_j \left(\calb^j \frac{\de S}{\de \calb^k}\right)\,.
\label{bfinalvar}
\eq

Lastly, we note that the velocity is determined via 
\bq
v^j =  \int_D d^3a\, \dot{q}^j \calj \de\left({r}-{q}\left(a,t\right)\right),
\eq
and we can use this to determine $\delta v$ in terms of $\de q$ and obtain the final Eulerian result. It is given by
\bqy
\label{vfinalvar}
v^j \p_k\frac{\de S}{\de v^j} &-& \p_k\left(v^j \frac{\de S}{\de v^j} \right) - \p_j \left(v^j \frac{\de S}{\de v^k}\right) \\ \nonumber
&-& \frac{\p}{\p t}\left(\frac{\de S}{\de v^k}\right).
\eqy
Together, equations (\ref{rhofinalvar}), (\ref{entropfinalvar}), (\ref{bfinalvar}) and (\ref{vfinalvar}) constitute the pieces that make up the equation of motion. Putting them all together, we have
\bqy
\label{equationofmotion1}
\calp \p_k\frac{\de S}{\de \calp} &+& \cals \p_k\frac{\de S}{\de \cals} - \p_k\left(\cals \frac{\de S}{\de \cals} \right) \\ \nonumber
+ \calb^j \p_k\frac{\de S}{\de \calb^j} &-& \p_j \left(\calb^j \frac{\de S}{\de \calb^k}\right) + v^j \p_k\frac{\de S}{\de v^j} \\ \nonumber
- \p_k\left(v^j \frac{\de S}{\de v^j} \right) &-& \p_j \left(v^j \frac{\de S}{\de v^k}\right) - \frac{\p}{\p t}\left(\frac{\de S}{\de v^k}\right) = 0. 
\eqy
It is evident   that \eqref{equationofmotion1} is fully Eulerian, since it  does not contain any Lagrangian pieces. Earlier, we'd mentioned that two different assumptions were made in building our model. Of these, we have used only the absence of the 1-forms in proving that our equation of motion is Eulerian. This assumption can \emph{also} be relaxed, and the ensuing result still remains the same. 

Now, we shall make use of the second assumption, namely the ansatz from (\ref{Theaction}) to recast (\ref{equationofmotion1}) into a more recognizable form. From the definition of the functional derivative, it can be shown that
\bq
\label{funcpartrel}
\frac{\de S}{\de \Psi} = \frac{\p \call}{\p \Psi} - \p_j \left(\frac{\p \call}{\p\left(\p_j \Psi\right)}\right),
\eq
where $\Psi$ represents any of the observables. This follows from the fact that $\call$ only involves the observables and their first-order spatial derivatives. Using this, one can rewrite (\ref{equationofmotion1}) as 
\bqy
\label{divfreeEOM}
-\frac{\p}{\p t}\left(\frac{\de S}{\de v^k}\right)&+&\partial_{j}\left[\delta^j_k \left(\mathcal{P}\frac{\delta S}{\delta\mathcal{P}}+\mathcal{B}^{j}\frac{\delta S}{\delta\mathcal{B}^{j}}-{\cal L}\right)\right] \\ \nonumber
&+&\partial_{j}\left[\frac{\partial{\cal L}}{\partial\left(\partial_{j}{\cal S}\right)}\left(\partial_{k}{\cal S}\right)+\frac{\partial{\cal L}}{\partial\left(\partial_{j}{\cal P}\right)}\left(\partial_{k}{\cal P}\right)\right] \\ \nonumber
&+&\partial_{j}\left[\frac{\partial{\cal L}}{\partial\left(\partial_{j}{\cal B}^{i}\right)}\left(\partial_{k}{\cal B}^{i}\right)+\frac{\partial{\cal L}}{\partial\left(\partial_{j}v^{i}\right)}\left(\partial_{k}v^{i}\right)\right] \\ \nonumber
&-& \partial_j \left[\calb^j \frac{\de S}{\de \calb^k} + v^j \frac{\de S}{\de v^k} + \dots \right]=0\, .
\eqy
It is important to clarify the notation employed in the above equation. The functional derivatives of $S$ are just the shorthand notation for the RHS of (\ref{funcpartrel}). Hence, it must be noted that the final expression only involves the partial derivatives of $\call$ with respect to the observables, and with respect to the spatial gradients of the observables. Lastly, we note that the ``$\dots$'' indicate that higher order derivatives of the observables can be included in the action (\ref{Theaction}), which induce higher order derivatives (and terms) in the above equation.

The equation of motion has been determined, and is given by (\ref{divfreeEOM}). Now, let us evaluate the dynamical equations for the  observables. From the Lagrange-Euler maps, one can use the procedure outlined in \cite{morrison09,ling} to obtain the corresponding dynamical equations. For $\calp$, we find that
\bq
\label{rhoevol}
\frac{\p \calp}{\p t} + \nabla \cdot \left(\calp v\right) = 0.
\eq
The dynamical equation for $\cals$ is found to be
\bq
\label{entropevol}
\frac{\p \cals}{\p t} + v\cdot\nabla \cals = 0,
\eq
and lastly, the evolution equation for $\calb$ is given by
\bq
\label{bevol}
\frac{\p \calb}{\p t} + \calb \left(\nabla \cdot v\right) - \left(\calb \cdot \nabla\right) v + \left(v \cdot \nabla\right) \calb = 0\,.
\eq


\section{Analysis of fluids, magnetofluids, and gyro fluids}
\label{Analysis}

In this section, we  use Noether's theorem in conjunction with (\ref{divfreeEOM}) to make some general statements about fluids and magnetofluids. Then, we shall specialize to the case of the gyrofluid and discuss it in greater detail. 

We work with the commonly used observables for fluid models, i.e.,  $\cals$ is replaced by $s$, $\calb$ by $B$ and $\calp$ by $\rho$, and  the action is decomposed into a part depending on   $\dot{q}$  and  one that does not: 
\bq
S[q] = \int \!dt\, \big(T[\dot{q}] - V[q]\big)\,.
\eq
It is important to note that there is no explicit $q$-dependence in our model. This arises from the fact that none of our Lagrange-Euler maps involve $q$ explicitly; instead, they involve only the derivatives of $q$ wrt $t$ and $a$. Since our action is fully Eulerianizable, it must involve the observables alone. None of the observables, on mapping back to their Lagrangian counterparts, involve $q$ explicitly.   In general, let us suppose that we can write $T[\dot{q}]$ as
\bqy
T[\dot{q}] &=& \int_D \!d^3a  \left(\calm_{0i} \dot{q}^i + \wp_{0ij} \dot{q}^i \dot{q}^j  + \calv_{0ijk}\dot{q}^i \dot{q}^j \dot{q}^k + \dots \right)
\nonumber\\
&&\hspace{-.75cm} + \int_D \!d^3r \left(\calm_{i} v^i + \wp_{ij} v^i v^j  + \calv_{ijk} v^i v^j v^k + \dots
\right)\,, 
\label{kinpot}
\eqy
where we have used the fact that our action is fully Eulerianizable. In other words, we require $\calm_{i}=\calm_{0i}/\calj$ and identical relations for $\wp$ and $\calv$ in order to ensure this property. Note that the RHS of this relation is evaluated at $a=q^{-1}(r,t)$ as always. 

We have not yet specified anything about the tensors $\calm$, $\wp$ and $\calv$. At this stage, we only know that they are functions of the observables and their spatial derivatives, i.e.,  they must possess the same form as $\call$ from (\ref{Theaction}), minus the dependence of $v$. Let us postulate further that these tensors are fully symmetric under the exchange of any pair of indices for the sake of simplicity. Since we know that our action is independent of $q$, the corresponding canonical momentum must be conserved. The canonical momentum is given by
\bq
\Pi_i = \calm_{0i} + 2 \wp_{0ij} \dot{q}^j + 3 \calv_{0ijk} \dot{q}^j \dot{q}^k + \dots,
\eq
since the tensors are symmetric. The Eulerian counterpart can be found from (\ref{Mcan3D}) by using the fact that $\calm_{i}=\calm_{0i}/\calj$ (and the same for the rest). It turns out to be
\bq
M^c_i = \calm_{i} + 2 \wp_{ij} v^j + 3 \calv_{ijk} v^j v^k + \dots.
\eq
This result can also be obtained from (\ref{divfreeEOM}), thereby serving as a consistency check. The first term in (\ref{divfreeEOM}), which is given by $-\frac{\p}{\p t}\left(\frac{\de S}{\de v^k}\right)$, reduces to $\partial{M^c_k}/\partial t$. As a result, our equation of motion becomes
\bq
\frac{\p M^c_k}{\p t} + \p_j T^j_k = 0,
\eq
which ensures that $M^c$ is conserved. The conservation of angular momentum is a much more trickier business. The sufficient condition for angular momentum conservation is that $T^j_k$ must be symmetric. Since we are dealing with a very general ansatz, it is not possible to determine \emph{a priori} whether our classes of models will conserve angular momentum in general. The quantities $\calm$, $\wp$, etc must be explicitly known in order to provide a definite answer. For the case of ideal hydrodynamics and magnetohydrodynamics, the tensor  $T^j_k$ is indeed symmetric.

Now, let us consider the simpler case wherein $\wp_{ij} = \frac{1}{2}\rho \delta_{ij}$. We define the kinetic momentum $M=\rho v$. We find that 
\bq
\label{McanMrel}
M^c_i = \calm_{i} + M_i + 3 \frac{\calv_{ijk}}{\rho^2} M^j M^k + \dots,
\eq
and we know that the LHS is conserved, i.e. $\frac{d}{dt}\int_D d^3r M^c$ is zero, provided we assume that the boundary terms vanish. Let us now consider the constraints under which the conservation of $M^c$ simplifies to the conservation of $M$. For starters, the first term on the RHS of (\ref{McanMrel}) must be expressible as the divergence of a tensor. Upon integration, it will then yield a boundary term which can be made to vanish. Hence, a sufficient condition for $M$ to be conserved is $\calm_i = \p_j \mathfrak{L}^j_i$ and $\calv_{ijk}=0$. 

We will now focus our attention on the model where the above constraints are satisfied. Let us choose to work with an action
\bq
\label{gyro3Daction}
S = S_{MHD} - \int_D d^3r dt\, \mathfrak{L}^j_i \p_j v^i,
\eq
where our set of observables are now $\rho$, $s$, $B$ and $v$. The quantity $S_{MHD}$ represents the ideal MHD action, whose explicit expression is known (see, e.g., \cite{morrison09}). From our preceding analysis, it is clear that both $M^c$ and $M$ are conserved for this model. It is also clear that this action satisfies the ansatz that we specified in (\ref{Theaction}). Furthermore, the ideal MHD action yields a symmetric momentum flux tensor, ensuring that $T^j_k$ is symmetric. Hence, the first term in (\ref{gyro3Daction}) also conserves angular momentum. 

As a result, we only need to investigate $\mathfrak{L}$ and the constraints that must be imposed upon it to ensure that $T^j_k$ is symmetric. Given that $\mathfrak{L}$ can only depend on $B$, $s$ and $\rho$ and their first order derivatives, there are still an infinite number of terms that can be generated.   It is evident of course that this system is too elaborate to permit further analysis. Hence, for starters, we shall assume that $\mathfrak{L}^j_i$ is symmetric and that it has the form
\bq
\label{specansatz}
\mathfrak{L}^j_i = \frac{1}{2}\left[\left(B^j B_i + B^i B_j\right) \alpha_I + \left(\delta^j_i + \delta^i_j\right) \alpha_{II} \right]\,, 
\eq
with   $\alpha_{I,II}$  only depending on $s$, $\rho$ and $|B|$.  We use (\ref{gyro3Daction}) and (\ref{specansatz}) in (\ref{divfreeEOM}). Rather than use brute force, we shall use some of the inherent symmetries of (\ref{divfreeEOM}). Note that the first line in (\ref{divfreeEOM}) contains terms that yield a symmetric contribution to $T^j_k$, since they are gradient terms, similar to the pressure. In the second line of (\ref{divfreeEOM}), there are no contributions since there are no density and entropy gradients. The same is also true for the first term on the third line of (\ref{divfreeEOM}), since (\ref{specansatz}) does not possess magnetic field gradients. As a result, we are left with only three terms of interest - the last three occurring in the LHS of (\ref{divfreeEOM}). Upon evaluation, we find that the resulting tensor is not symmetric, and the ansatz (\ref{specansatz}) does not possess angular momentum conservation.

Now, let us suppose that we consider the hydrodynamic case where $B$ is absent. We shall consider the case where gradients wrt $s$ and $\rho$ are absent, and the gyroscopic term is of the form (\ref{gyro3Daction}).  The condition for angular momentum conservation becomes particularly simple, since the tensor
\bq
\mathfrak{L}^j_i \left(\partial_{k}v^{i}\right) + v^j \p_i \left(\mathfrak{L}^i_k\right),
\eq
must be symmetric. 


\section{An illustration of the formalism}
\label{rotMHD}

To demonstrate the utility of the formalism developed in this paper,  we  now consider a simple illustration that demonstrates how an additional attribute can be added to ideal HD.  The attribute we add represents an internal degree of freedom, an intrinsic rotational (angular) velocity, attached to each fluid element.  Given the new set of observables, we  immediately use (\ref{divfreeEOM}) to compute the corresponding equation of motion.

There are many physical situations where an internal angular velocity or momentum is appropriate, because such microscopic behavior influences the macroscopic dynamics.  One example is the effect of finite Larmor radius gyration of charged particles in a magnetic field while another occurs in the  the theory of nematodynamics.  We consider the latter which applies to  liquid crystals that are  modeled as a fluid composed of rigid rods. These rods are endowed with a preferred direction, called the director, and an intrinsic angular momentum. The relevant dynamics for this system are described in \cite{st05} (with details in the classic works of \cite{lu70,fo71,mar72,st74}).  A simplified  limit of this work where phenomenological dissipative relaxing is removed (their parameter $\gamma^{-1}\rightarrow 0$) results in a reduction to a single variable $\Omega_\parallel$, an angular velocity proportional to the intrinsic angular momentum parallel to the now constant director.  The new variable $\Omega_\parallel$  is advected by the fluid velocity field, and thus behaves as  a zero form. We shall work with this subcase henceforth.

One must now construct an appropriate Lagrange-Euler map for our attribute-observable pair, denoted by 
$\Omega_{0\parallel}$ and $\Omega_\parallel$, respectively. Since we have noted that $\Omega_\parallel$ is advected, this corresponds to $\Omega_\parallel=\Omega_{0\parallel}$, with the RHS evaluated at $a=q^{-1}(r,t)$. The advection equation is given by
\bq
\label{omevol}
\frac{\p \Omega_\parallel}{\p t} + v^j \p_j \Omega_\parallel = 0,
\eq
and the similarity to the entropy is self-evident. One can now construct an angular momentum  density, $l^2\omega_d := \rho\,  l^2\Omega_\parallel$,  that behaves as a three-form. Here the quantity $l$ represents the radius of gyration with  $l^2$  being  interpreted as  the moment of  inertia per unit mass.  Its  governing equation is
\bq
\frac{\p \omega_d}{\p t} + \p_j \left(v^j \omega_d\right) = 0,
\eq
and the relationship between $\omega_d$ and its corresponding attribute $\omega_{0d}$ is $\omega_d=\omega_{0d}/\calj$ with the RHS  evaluated at  $a=q^{-1}(r,t)$.

By  analogy with classical (discrete) mechanics, we propose the continuum rotational kinetic energy functional, 
\bq
\label{rotkin}
K_{rot}:= \int_D \!d^3a\, \frac{1}{2} \rh_0 l^2 \Omega_{0\parallel}^2 = \int_D \!d^3a\, l^2 \frac{\omega_{0d}^2}{2\rh_0}\,.
\eq
 It is easily verified that the above functional satisfies the Eulerian closure principle, with its counterpart given by  $l^2 \omega_d^2/(2\rh)$. Since (\ref{rotkin}) is entirely independent of $q$, it serves as a Lagrangian invariant, and does not enter the equation of motion. This can also be verified by taking the Eulerian counterpart and substituting it into (\ref{divfreeEOM}). Despite its absence in the momentum equation of motion, it is instructive to see how the rotational and translational kinetic energies stack up against each other. In order to compute the rotational energy, we assume that our fluid particles can be modeled as molecules. In such a scenario, we find that the ratio reduces to 
\bq
\frac{l^2 \Omega_\parallel^2}{v^2} \sim \frac{\Theta}{T},
\eq
where $\Theta$ denotes the rotational temperature \cite{rei98}. We have assumed that $v$ is characterized by the thermal velocity, and $l \Omega_\parallel$ by the rotational temperature. In general, it is evident that this ratio is extremely small for hot fluids, such as the ones observed in fusion reactors or in stars. However, there exist environments in nature, such as giant molecular clouds which possess temperatures of a few tens of Kelvin \cite{shu87,mck07}. They are comprised of molecular hydrogen, whose rotational temperature is known to be around 88 K \cite{rei98}. As a result, we see that the two energies are comparable in this regime and there remains an outside possibility that such effects might be of importance.

We have earlier mentioned that we study a subcase of \cite{st05} where coupling terms involving $v$ and $\Omega_\parallel$ are non-existent. Now, let us  add  a simple term of the form $l^2  \omega_d C_{i}^k \p_k v^i$ to  the action, where $C_{i}^k$ is a tensor with constant coefficients. Then, the  full action is given by
\bq
\label{rotHD}
S:= S_{HD} + \int_D d^3r dt\, l^2 \omega_d C_{i}^k \p_k v^i,
\eq
where $S_{HD}$ represents the ideal HD action. The new term can be interpreted as follows. Integration by parts casts it in the form of $v\cdot \nabla \times L_{int}$, if one associates the tensor $C^k_{i}$ with the three-dimensional Levi-Civita tensor (with one of the indices fixed to be $\hat{z}$, the director direction) and the term $L_{int}$ with the intrinsic angular momentum density. By inspection, it is found that $L_{int} = \left(l^2 \omega_d\right) \hat{z} = \left(\rho l^2\right) \Omega_\parallel \hat{z}$ and it is evidently the product of the moment of inertia (per unit volume) and the angular velocity. This term was not constructed at random - it corresponds to the analogue of 2D gyroviscous MHD studied in \cite{ling}. In gyroviscous MHD, the particles undergo Larmor gyration as a result of the magnetic field, behaving as though they were indeed endowed with an intrinsic angular velocity (and angular momentum). The corresponding equation of motion is given by
\bqy \label{spinEOM}
\frac{\p \left(\rho v_k\right) }{\p t} &+& \p_j \left[\left(C_{k}^j v^i - C_{k}^i v^j\right) \p_i \omega_d \right] \\ \nonumber
&+& \p_j \left[ \omega_d \left(C_{k}^j \p_i v^i - C_{i}^j \p_k v^i\right) \right] + \dots = 0,
\eqy
where the  ``$\dots$'' indicate that this corresponds to the ideal HD equation of motion. The four additional terms involve gradients with respect to the velocity (or angular velocity), and they serve as the \emph{de facto} viscosity tensor. If we assume that the fluid has the property that $\omega_d = \mathrm{const}$, the two terms in the first line of (\ref{spinEOM}) vanish identically. However, the next two terms are still present, which changes the ideal MHD momentum flux. With this special choice of $\omega_d$, the similarities with the orthodox viscous tensor are striking - there are terms involving $\partial_k v^i$ and the divergence $\partial_i v^i$, and the coefficients in front of these terms correspond to the dynamic and bulk viscosities respectively. Thus, we see that the angular momentum  fluid, with some minor restrictions, mirrors the conventional viscous fluid. In general, $\omega_d$ depends on time, and hence one can interpret (\ref{spinEOM}) as comprising of time-dependent viscosities, thereby representing a theory of non-Newtonian fluids \cite{tru65,ast74}. The importance of such fluids in biological systems is well-documented \cite{das06,ber08}.
The action (\ref{rotHD}) conserves energy and linear momentum $\rho v$, 
but not the angular momentum $r \times (\rho v)$.

In our discussion here, we have built a theory of fluids with intrinsic angular momentum by incorporating the rotational kinetic energy and gyroviscous terms. This illustrative model corresponds to a simplified version of \cite{st05} for nematic effects in liquid crystals, but  with additional effects incorporated, and was presented to demonstrate how to build models from scratch.  Clearly the  nondissipative parts of more complete models can be built in this manner, and potential applications in a variety of fields, e.g., nematics,  micromorphic systems \cite{for09}, and plasma physics, come to mind.   


\section{Conclusion}
\label{Conclusion}

In this paper, we have presented a general class of actions, described by the ansatz (\ref{Theaction}). This class includes ideal MHD, symmetric MHD, reduced MHD, gyroviscous MHD and their HD equivalents. For this class of actions, we have shown that an Eulerian action gives rise to Eulerian equation(s) of motion, and presented the explicit form for the latter. By making use of this result, we present a general analysis of the conditions under which momentum (and angular momentum) is conserved. Lastly, as an illustrative application of this formalism, we used it to study HD models where the fluid particles possess an intrinsic angular velocity (and angular momentum). It was shown that these models behaved akin to viscous HD models, but conserved energy and linear momentum, but not the angular momentum. These models may prove to be of significance in certain astrophysical environments, and in studying nematic and biological systems.

One of the chief advantages of this approach stems from the potential application to the two-fluid model action by incorporating gyroviscous effects. Such a procedure would amount to an inherently consistent, first-principles derivation of a two-fluid gyroviscous tensor, which can then be compared against the Braginskii gyroviscous tensor \cite{braginskii65}. Similarly, the formalism developed in this paper can be extended to include kinetic and gyrokinetic theories, which can then be analyzed to study wide ranging plasma and astrophysical phenomena. It is also possible to use the models from 
Sec.~\ref{rotMHD} to study their implications for momentum and angular momentum transport in astrophysical contexts.  In the future \cite{pjmMW14}, we shall use the HAP formalism presented herein to derive gyroviscous tensors, and to incorporate an anisotropic pressure into the equation(s) of motion.


\section*{Acknowledgment}
\noindent   Supported by U.S. Dept.\ of Energy Contract \# DE-FG05-80ET-53088.



\begin{thebibliography}{58}
\bibitem{EP} 
Such induced variations have a long history.  For example, finite-dimensional analogs of these variations were used by Poincar\`e and Hamel, while fluid versions were used in the MHD context in \cite{FR60,newcomb62,amp2a}. In the context of kinetic theory they appeared in \cite{cendra}, where they were called Euler-Poincar\'e variations,   a variety of  complex fluids were treated in \cite{gbr}, and hybrid plasma models in \cite{tronci2}.
\bibitem{dual}  In MHD one usually considers the dual to the two-form, the $B$-field, but there is some subtlety here.   It was shown in \cite{morrison82} that MHD is a Hamiltonian theory even when $\nabla\cdot B\neq 0$.   The Hamiltonian equations are then interpreted as  the Lie dragging of a vector density of rank unity.
\end{thebibliography}
\end{document}